\documentclass[aps,prb,twocolumn,amsmath,amssymb,superscriptaddress,showpacs]{revtex4-1}

\newcommand{\be}{\begin{equation}}
\newcommand{\ee}{\end{equation}}
\newcommand{\bea}{\begin{eqnarray}}
\newcommand{\eea}{\end{eqnarray}}
\usepackage[utf8]{inputenc}
\usepackage{amsfonts}
\usepackage{amsthm}
\usepackage{fontenc}
\usepackage{graphicx}
\usepackage{textcomp}
\usepackage{epstopdf}
\usepackage{braket}
\usepackage{mathtools}
\usepackage{dcolumn}
\usepackage{bm}
\usepackage{epsfig}
\usepackage[usenames,dvipsnames]{color}
\usepackage{booktabs}
\usepackage{units}
\begin{document}
\newcommand{\abs}[1]{\lvert#1\rvert}
\title{Gaussian deformations in graphene ribbons: flowers and confinement}
\author{R. Carrillo-Bastos}
\affiliation{Centro de Investigaci\'{o}n Cient\'{i}fica y de
Educaci\'{o}n Superior de Ensenada, Apado. Postal 360, 22800
Ensenada, Baja California, M\'{e}xico} \affiliation{Universidad
Nacional Aut\'{o}noma de M\'{e}xico, Apdo. Postal 14, 22800
Ensenada, Baja California, M\'{e}xico } \affiliation{Ohio
University, Athens, Ohio 45701-2979, USA}
\email{nomarcb@hotmail.com}
\author{D. Faria}
\affiliation{Universidade Federal Fluminense, Niter\'{o}i, Avenida
Litor\^{a}nea sn, 24210-340 RJ, Brasil}
\author{A. Latg\'{e}}
\affiliation{Universidade Federal Fluminense, Niter\'{o}i, Avenida
Litor\^{a}nea sn, 24210-340 RJ, Brasil}
\author{F. Mireles}
\affiliation{Universidad Nacional
Aut\'{o}noma de M\'{e}xico, Apdo. Postal 14, 22800 Ensenada,
Baja California, M\'{e}xico }
\author{N. Sandler}
\affiliation{Ohio University, Athens, Ohio 45701-2979, USA}

\begin{abstract}
The coupling of geometrical and electronic properties is a
promising venue to engineer conduction properties in graphene.
Confinement added to strain allows for interplay of different
transport mechanisms with potential device applications. To
investigate strain signatures on transport in confined geometries,
we focus on graphene nanoribbons (GNR) with circularly symmetric
deformations. In particular, we study GNR with an inhomogeneous,
out of plane Gaussian deformation, connected to reservoirs. We
observe an enhancement of the density of states in the deformed
region, accompanied with a decrease in the conductance, signaling
the presence of confined states. The local density of states
exhibits a six-fold symmetric structure with an oscillating
sub-lattice occupation asymmetry, that persist for a wide range of
energy and model parameters.
\end{abstract}


\maketitle
Graphene nanoribbons (GNR)
constitute a viable way to exploit the extraordinary electronic
transport properties of graphene\cite{GrafRMP}. The rich
combination of peculiar properties due to confinement and
potential technological applications have guided research on
nanoribbons since the pioneering studies by Nakada et al
\cite{Nakada}. With improved control of growth and manipulation
techniques of graphene flakes and carbon nanotubes\cite{Ma,Tao},
experimental studies focused on different aspects of the physics
displayed in these reduced geometries. Original works confirmed
the appearance of gaps due to confinement\cite{Han-Kim,Avouris},
and later experiments demonstrated the stability of zigzag
terminated structures\cite{Jia,Girit}. Further studies focused on
issues such as tailored edge terminations\cite{Chuvilin,Zhang},
atomic scale control of electric contacts\cite{vanderLit}, and
transport properties at high biases\cite{Han-Kim}. Furthermore, it
was recently reported that ribbons grown epitaxially on SiC can
stand ballistic transport on length scales greater than $10$
$\mu$m\cite{NRNature}, a finding very relevant for electronic
applications. The fast pace of experimental studies is stimulated
-and accompanied- by a vast amount of theoretical work predicting
a wide variety of phenomena from localized magnetic properties at
the edges, to exotic topological phases\cite{Yazyev,JuanJose}.
Recently, studies have begun to address the effect of strain
in transport properties of confined systems and ribbon junctions.

Strain in graphene has been the topic of a large number of
theoretical
works\cite{Fernando1,GuineaGeim,Vitor1,Low,Fernando,MFV,Barraza1,Barraza2,Barraza3,Gerardo}
aimed at understanding the effects of controlled deformations on
electronic properties. As charge carriers near the neutrality
point behave as massless Dirac particles moving on a deformed
lattice, many aspects of fundamental physics involved in the
dynamics of such system can be studied in great detail on current
settings. Experimental works have analyzed different aspects of
strain on graphene: from the initial measurements of its intrinsic
strength and elastic properties\cite{Hone} to the more recent
identification of pseudo-Landau levels (LL) associated to gigantic
pseudo magnetic fields produced in highly strained
samples\cite{Bubbles}. These achievements are accompanied by the
development of devices such as strain based graphene
sensors\cite{Ahn} and piezoelectrics, among others, giving rise to
the nascent field of straintronics\cite{Ong}.

\begin{figure}
\includegraphics[scale=0.25]{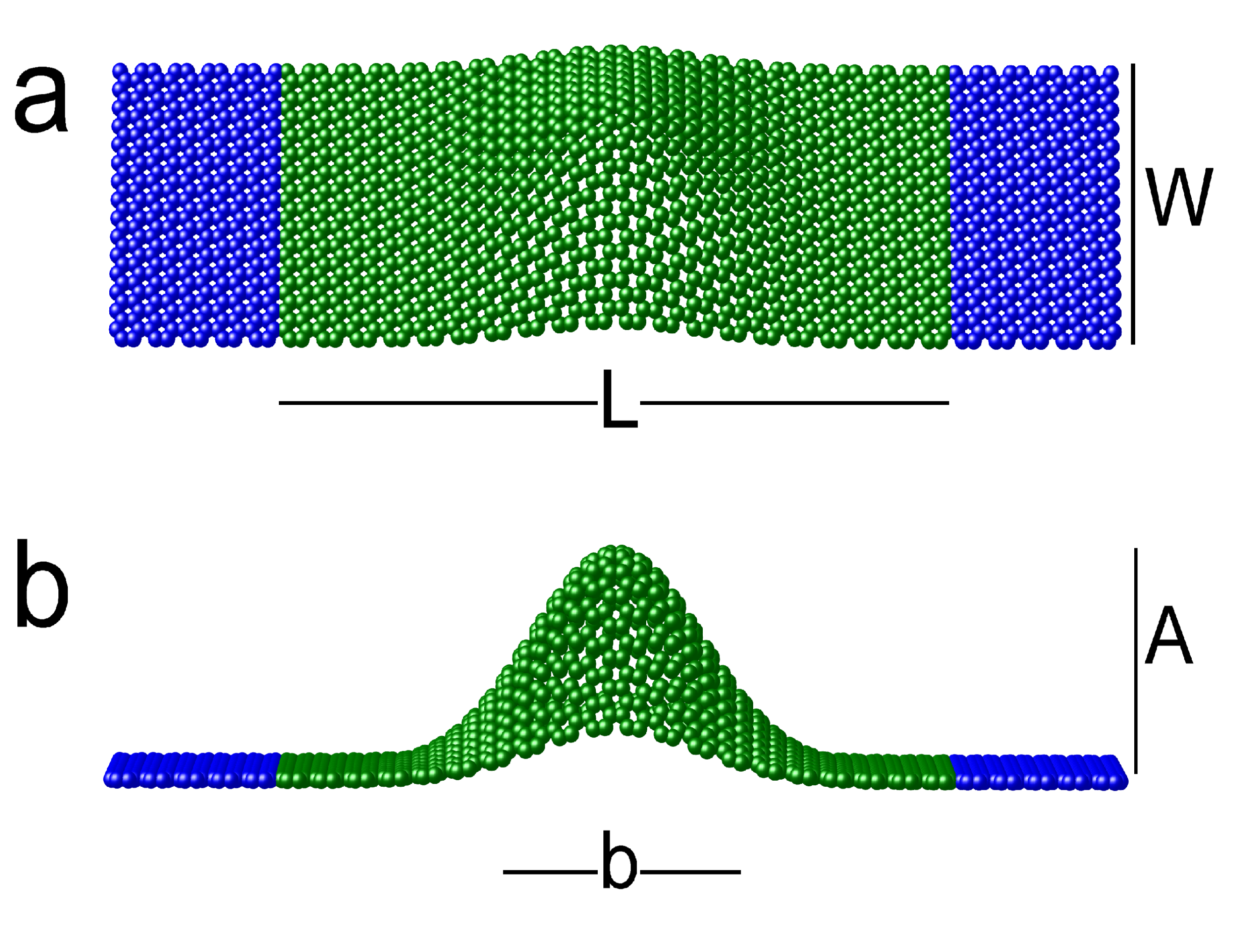}
\caption{(Color online) Schematic representation of deformed GNR
(width $W$ and length $L$) connected to leads with a Gaussian
deformation (amplitude $A$ and dispersion $b$). \label{Fig1}}
\end{figure}

Although a good degree of understanding of homogeneous strain has
been achieved in recent years, the role of non-uniform strain, and
in particular, in confined open geometries, remains still
unexplored. The purpose of this paper is to provide insight into
this issue by studying equilibrium and transport properties of
strained nanoribbons with armchair and zigzag edges connected to
reservoirs. Most of the theoretical work on transport in deformed
graphene has focused on uniaxial
strain\cite{PRB80.045401,PRB88.195416,Falko2,PRB81.161402,PRL101.226804},
with centro-symmetric deformations being analyzed in closed
geometries\cite{Moldovan,Daiara,Wakker,PRB84.081401} or in open
systems within the Born approximation\cite{JAP112.073710}. We
focus here on strain produced by a centro-symmetric Gaussian (out
of plane) deformation located at the center of the nanoribbon as
shown in Fig.~\ref{Fig1}. These deformations can serve as a model
for a load in a membrane\cite{Barraza1}, ripples in free standing
graphene\cite{Moneesh} or gaussian patterns in
substrates\cite{PRB83.165405}. They have been already produced on
suitable substrates\cite{BubblesControl} and also with STM
methods\cite{Morgenstern,Klimov}. Two main results emerge from
this study: i) these deformations confine states within the
ribbon, with the consequent decrease of the conductance and
corresponding peaks appearing in the associated density of states
(DOS); ii) the local density of states (LDOS) exhibits a $6-$fold
symmetry pattern that we refer to as the 'flower', with sublattice
polarization in each sector (or 'petal'). These results, independent of the crystalline orientation, exhibit
new interesting features discussed below, and are analogous to the
proposed Dirac fermion confinement with real magnetic
fields\cite{Egger}. Our study focuses on transport mechanisms
different from LL-assisted tunneling\cite{Zenan} that has been  proposed to explain recent experimental\cite{Bubbles} and theoretical results obtained for
strained ribbons\cite{Falko1,Falko2}.

\noindent{\it Model for GNR with Gaussian Deformation}:
We consider a nanoribbon with $N_x$ ($N_y$) sites on the horizontal (vertical)
direction, connected to infinite graphene leads (see Fig.~\ref{Fig1})
modeled by the tight-binding Hamiltonian:
\be
\label{hamiltonian}
 H=\sum\limits_{<i,j>} t_{ij}c_i^\dagger c_j + \sum\limits_{<i,k_l>} t_{0}c_i^\dagger c_{k_l} + \sum\limits_{<i,k_r>} t_{0}c_i^\dagger
 c_{k_r},
\ee
Here, the first term refers to the central (deformed) system,
while the second and third terms describe the connection to the
reservoirs, with the indices $k_l, k_r$ running over the sites of
the left and right leads. $c_i^\dagger$ ($c_i$) is the creation
(annihilation) field operator in the $i$-th site, $t_{ij}$ is the
nearest-neighbor hopping energy and we take $t_{0}= -2.8 $ eV as
the hopping parameter in the absence of deformation. The strain
introduced by the Gaussian deformation modifies $t_{ij}$  as
$t_{ij} = t_{0}\Delta_{ij}$ with $\Delta_{ij}=e^{-\beta \left(
\nicefrac{l_{ij}}{a}-1\right)}$.

The interatomic distance in unstrained graphene is $a=1.42\
\AA$, and the coefficient $\beta=\left|\frac{\partial\log
t_o}{\partial\log a}\right|=3.37$. The distance
$l_{ij}=\frac{1}{a}\left(a^{2}+\varepsilon_{xx}x_{ij}^2+\varepsilon_{yy}y_{ij}^2+2\varepsilon_{xy}x_{ij}y_{ij}
\right)$ is given by the strain tensor
$\varepsilon_{\mu\nu}=\frac{1}{2}\left(\partial_\nu
u_\mu+\partial_\mu u_\nu+\partial_\mu h \partial_\nu h\right)$,
characterized by the in- and out-plane deformation, $u_\nu$ and
$h$, respectively\cite{Landau}. The out-of-plane deformation,
\be
h\left( x_{i}, y_{i}\right) =A
e^{-\frac{(x_{i}-x_{0})^2+(y_{i}-y_{0})^2}{b^2}},
\label{gaussian}
\ee
has center at $[x_{0}; y_{0}]$ (we use $\left(x_0=L/2,y_0=W/2
\right)$ for the center of the ribbon), and $A$ and $b$ describe
its amplitude and width respectively. The hopping modification can
be understood as a gauge field \cite{Ando}. For a Gaussian
deformation this field has a three-fold spatial distribution with
different profiles for zigzag and armchair crystal
directions\cite{Daiara, Moldovan}. Notice that the bump also
produces a deformation potential, akin to a local chemical
potential\cite{Ando}, whose effects have not been included in the
results showed below. Consequences of its presence are discussed
in detail in the Supplemental Materials\cite{Supplementary} where
it is shown that due to its axial symmetry, it does not affect the
main findings of this paper. Eq.~\ref{hamiltonian} is used to
obtain the retarded Green's function by recursive methods.
Self-energies $\Sigma_{r,l}$ associated to the leads, are
calculated by standard decimation methods. Finally, the
conductance is calculated via the Landauer formula and Fisher-Lee
relation\cite{Mucciolo}.

\begin{figure}
\includegraphics[scale=0.40]{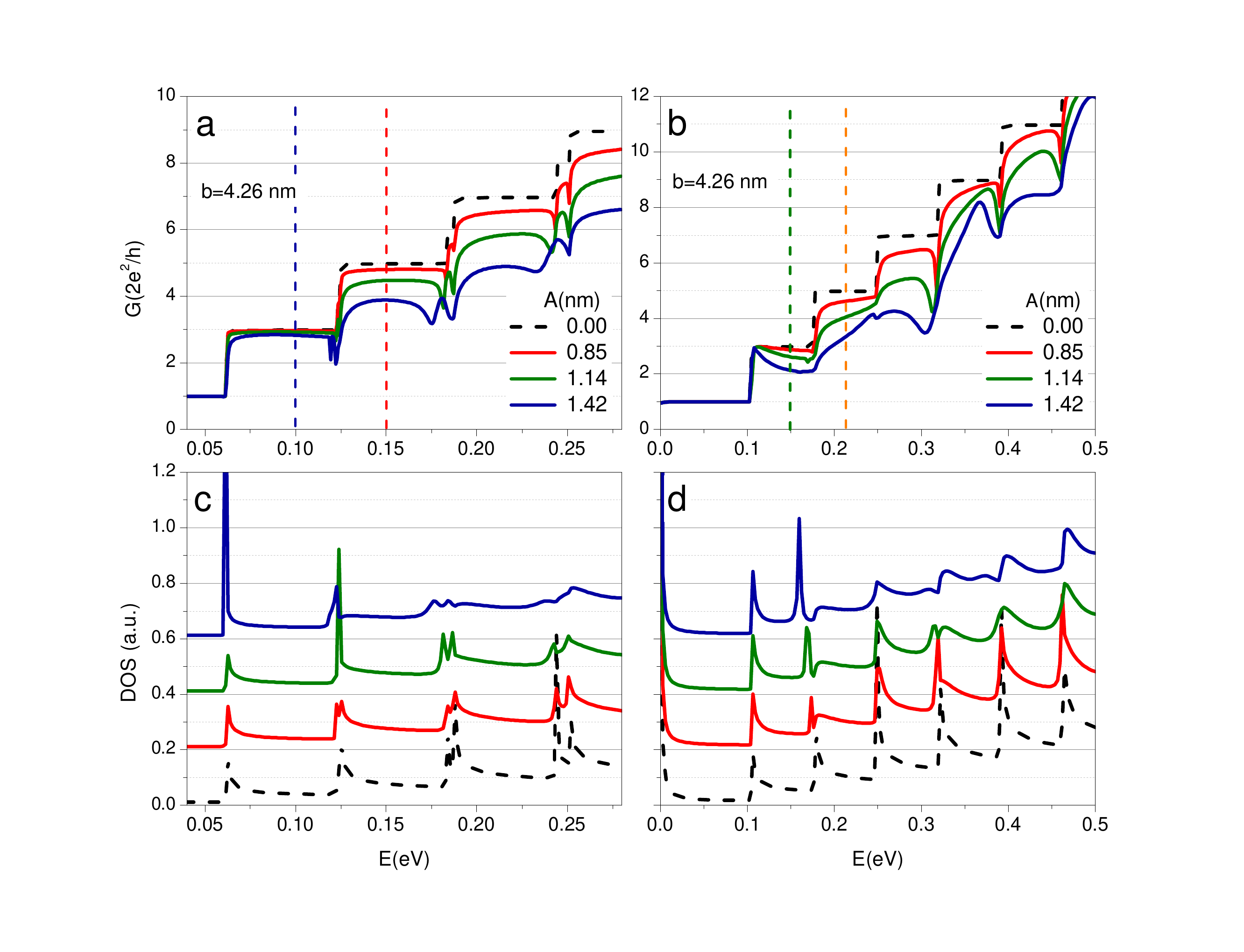}
\caption {(Color online) Conductance and DOS for: AGNR [a) and c)], and ZGNR
[b) and d)] for different values of $A$, and fixed width $b=4.3
nm$, corresponding to strains of $4, 7$ and $11\%$ respectively.
Black dashed lines corresponds to the non-deformed ribbon. Data
for deformed DOS has been shifted for clarity. Vertical lines
indicate energies values at which LDOS plots are shown in
Fig.~\ref{Fig4}. \label{Fig2}}
\end{figure}

\noindent{\it Conductance and DOS:} The conductance and DOS for
strained ribbons with armchair (AGNR) and zigzag (ZGNR)
terminations are shown in Fig.~\ref{Fig2} for deformations with
varying amplitude $A$ and fixed width $b$. In both cases the
position of the deformation is at the center of one hexagonal
cell. The data is shown for AGNR with $L=30.7$ nm and $W=30.0$ nm
($288\times245$ atomic sites) and for ZGNR with $L=27.4$ nm and
$W= 25.8$ nm ($224\times244$ atomic sites). Similar results were
observed with different ribbon sizes and positions of the Gaussian
center (within a radius of $\sim 0.2$ nm). For all panels, the
dashed (black) lines correspond to results in the absence of the
deformation and continuous (color online) lines to different
values of A.

Conductance results are shown in panels a) and b) for AGNR and
ZGNR, respectively. Both ribbons are metallic and the conductance
exhibits the standard stepwise behavior for the unstrained case
(black dashed). For both terminations, the zero-plateau is not
modified by the Gaussian deformation, in contrast with results
obtained with uniaxial in-plane strained
junctions\cite{PRB88.195416}. As $A$ increases, the value of
 the conductance decreases for non-zero plateaus. Note that the conductance
 for ZGNR and AGNR ribbons exhibit different profiles. These differences may be caused by
the distinct orientations of the pseudo magnetic
field space distributions with respect to the position of the
leads. These distributions are  $90^{\circ}$ rotated with respect to each other resulting on different scattering cross sections
as shown by perturbation theory calculations on the continuum
model\cite{JAP112.073710}.  A common feature for both ribbons is the appearance of
pronounced minima at the step-to-step transition, which have been observed in other systems,
and are associated with inter-band mixing favored by the presence of perturbations\cite{PhysRevB.41.10354}.

Lower panels c) and d) show results for the corresponding DOS. The
DOS curve for ZGNR shows the peak at zero energy corresponding to
edge states that remains largely unaffected by the deformations
from $4\%$ up to a level of $11\%$ strain. As the deformation is
turned on, for both terminations, sharp peaks appear at lower energies followed by local minima. These minima are followed by raising features precisely at the energy values corresponding to
the van Hove singularities in the absence of
deformation. Thus, peaks in the undeformed system shift
spectral weight lower energy peaks. These new
peaks --in contrast to the original ones-- are symmetric, fact
more evident for ZGNR (see for example the
third peak. This effect is accompanied by a decrease in height of
higher energy peaks and a slow smoothening of the DOS. Notice that states in the newly formed low-energy DOS peaks,
(produced by inhomogeneous pseudo-magnetic field), do not generate
additional contributions to the conductance. This indicates an
incipient localization at the deformation region, in contrast with
previous studies where extended regions with constant
pseudo-magnetic field generate pseudo LLs available for
tunnelling assisted transport\cite{Falko2,Zenan}.

\begin{figure}
\includegraphics[scale=0.40]{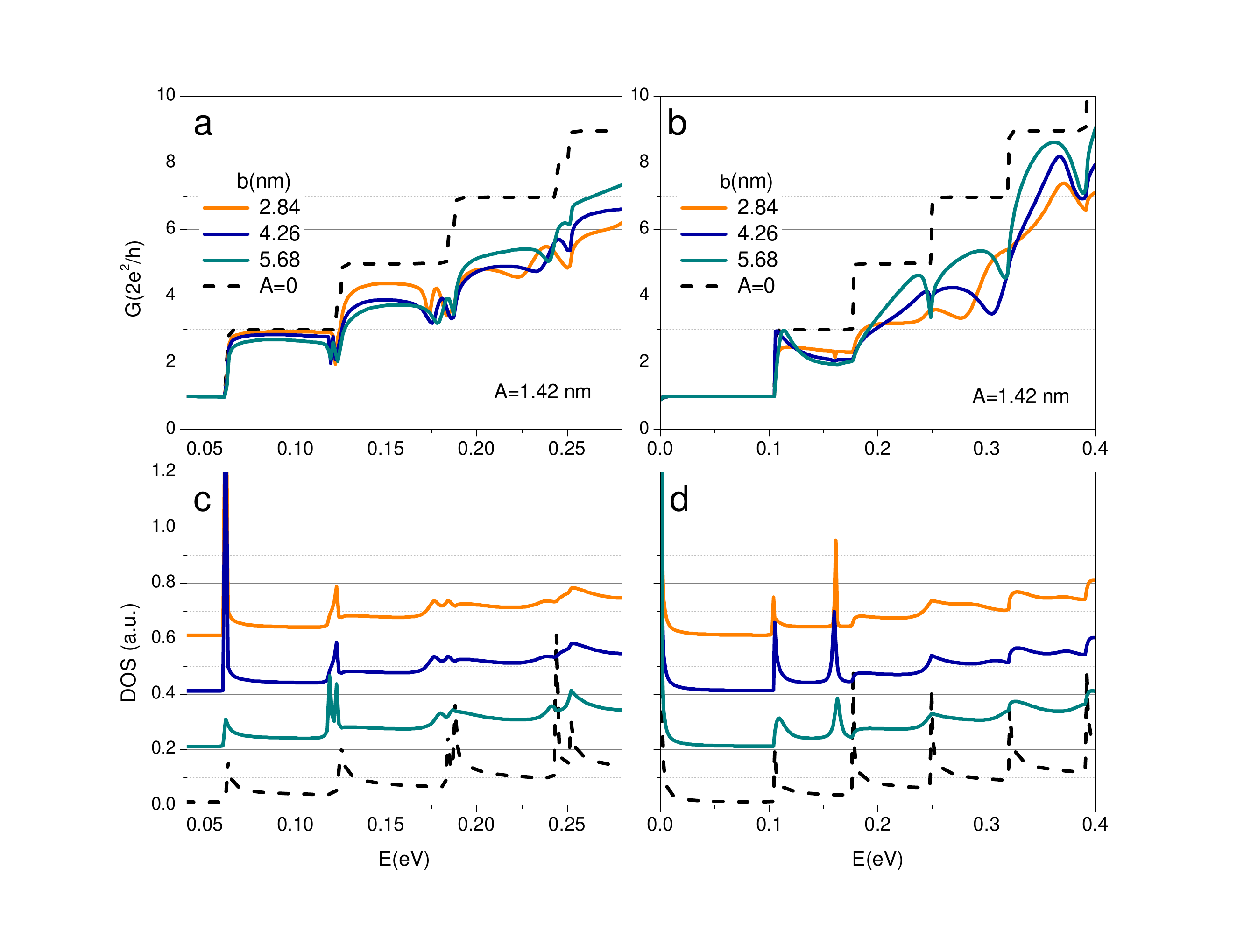}
\caption {(Color online) Conductance and DOS for: AGNR [a) and
c)], and ZGNR [b) and d)] for AGNR, different values of $b$, and
fixed amplitude $A = 1.42 nm$, corresponding to strains of $7$,
 $11$,  and $25$,  respectively. The (black) dashed line corresponds
to the non-deformed ribbon. Data for deformed DOS has been shifted
for clarity.} \label{Fig3}
\end{figure}

Fig.~\ref{Fig3} shows similar results for a deformation with
constant amplitude $A$ and variable width $b$. For both ribbons
terminations, an increase in the curvature of the deformation (decreasing the value of $b$)
results in a deterioration of the conductance and confined states.
We find that the energy of the newly confined level decreases
quadratically with the aspect ratio $(A/b)$, a result predicted in
the continuum description (Dirac) by perturbation theory and
confirmed by scattering calculations\cite{Martin}.

\begin{figure}
\includegraphics[scale=0.215]{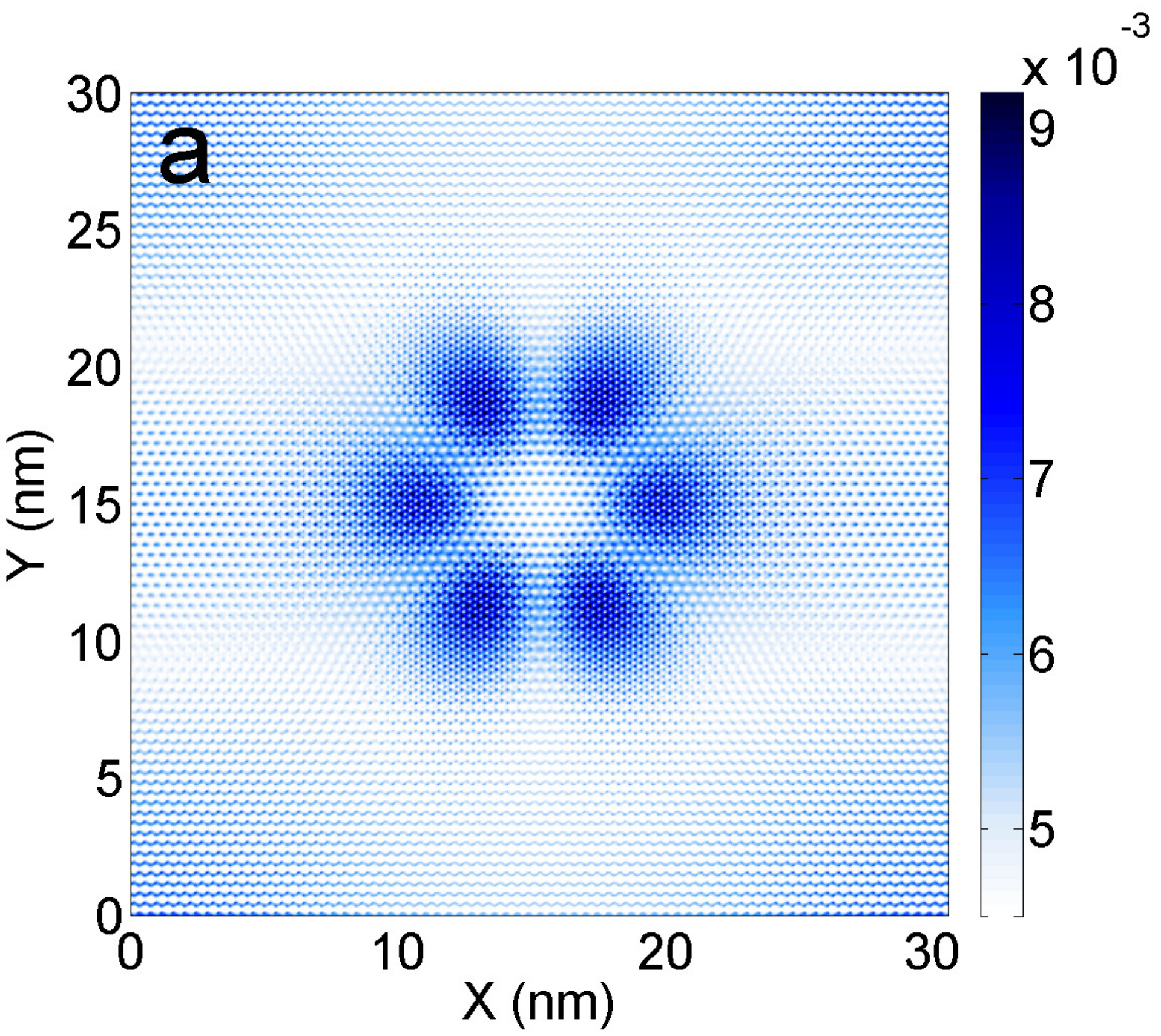}
\includegraphics[scale=0.215]{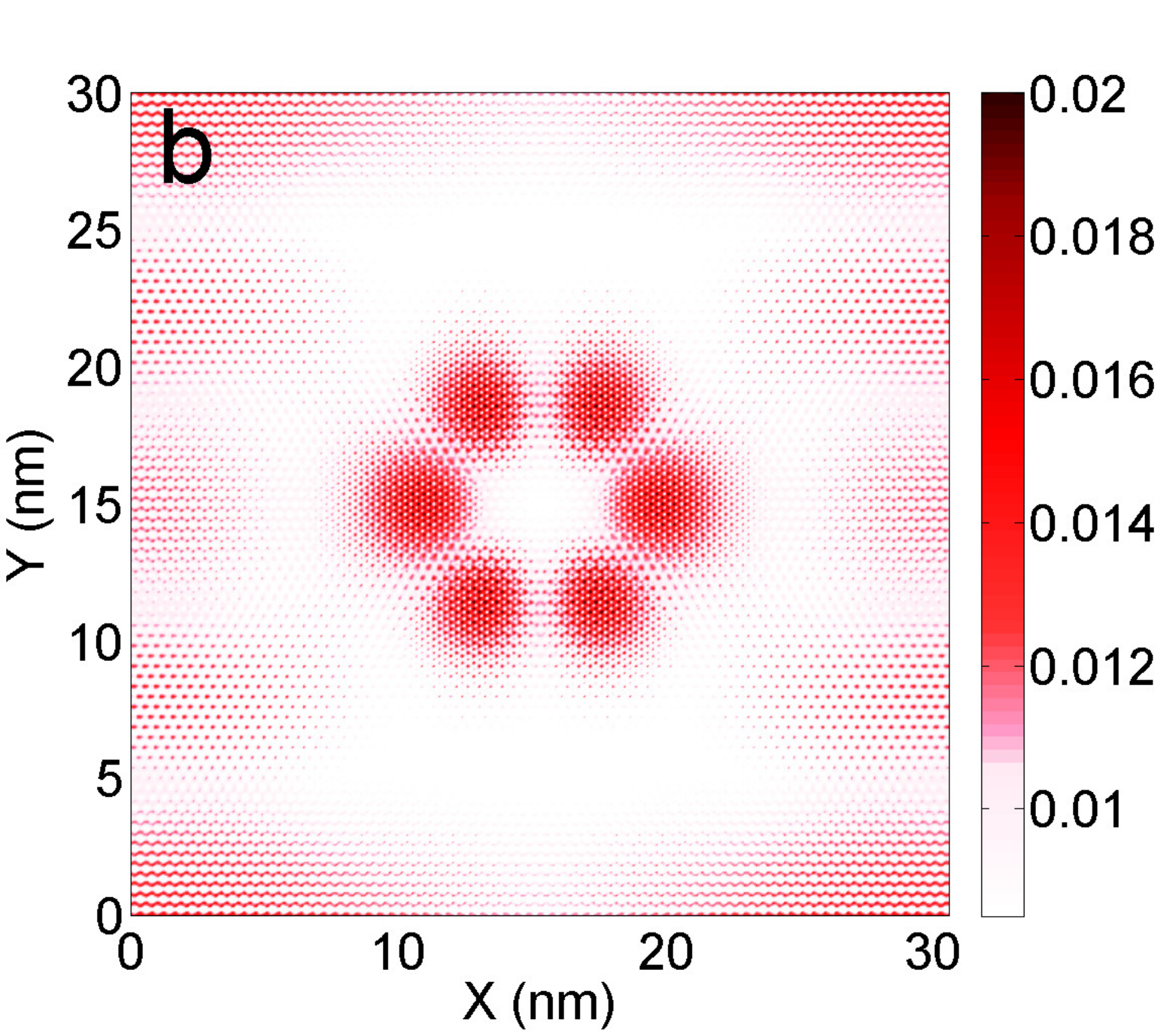}
\includegraphics[scale=0.215]{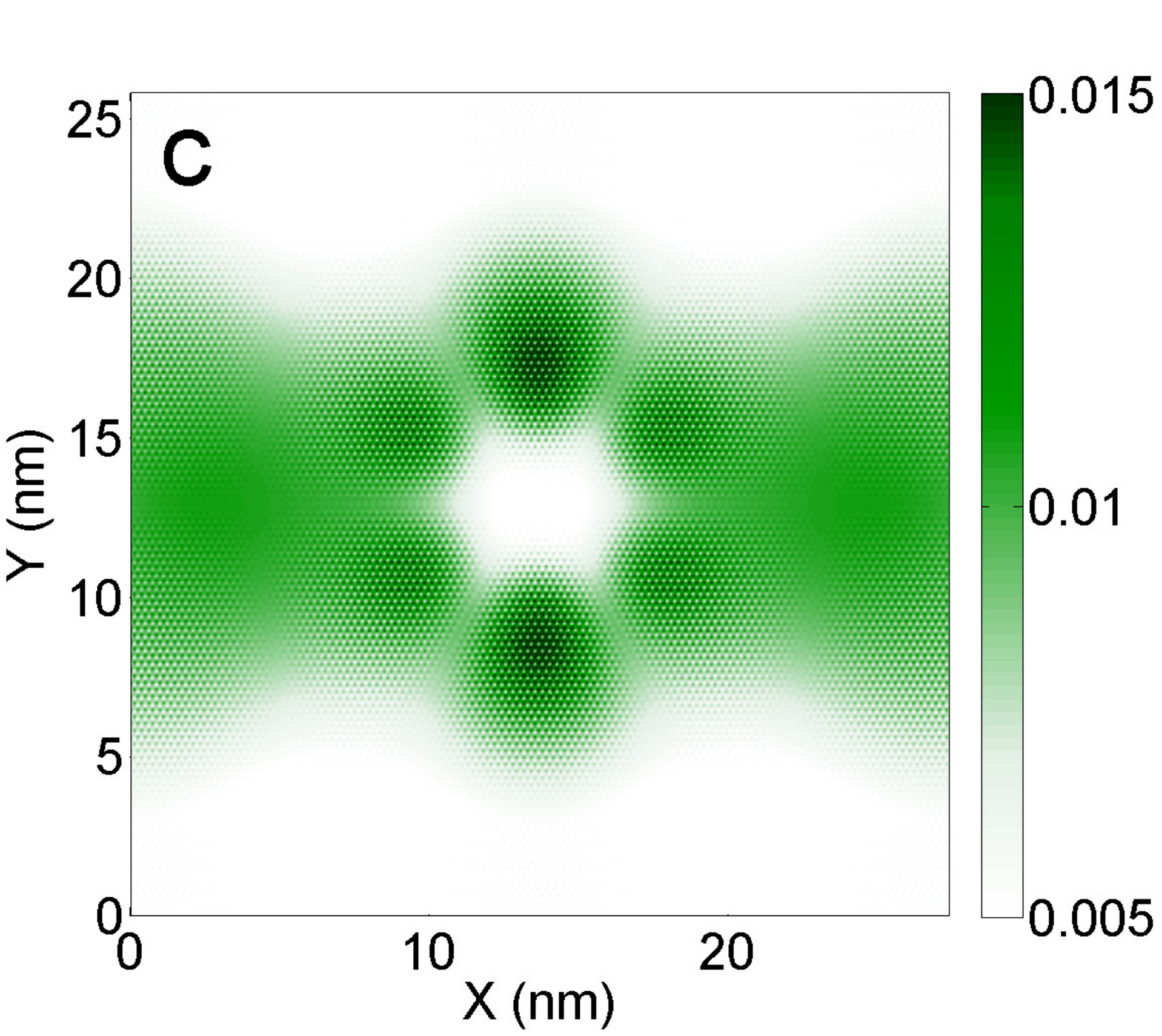}
\includegraphics[scale=0.215]{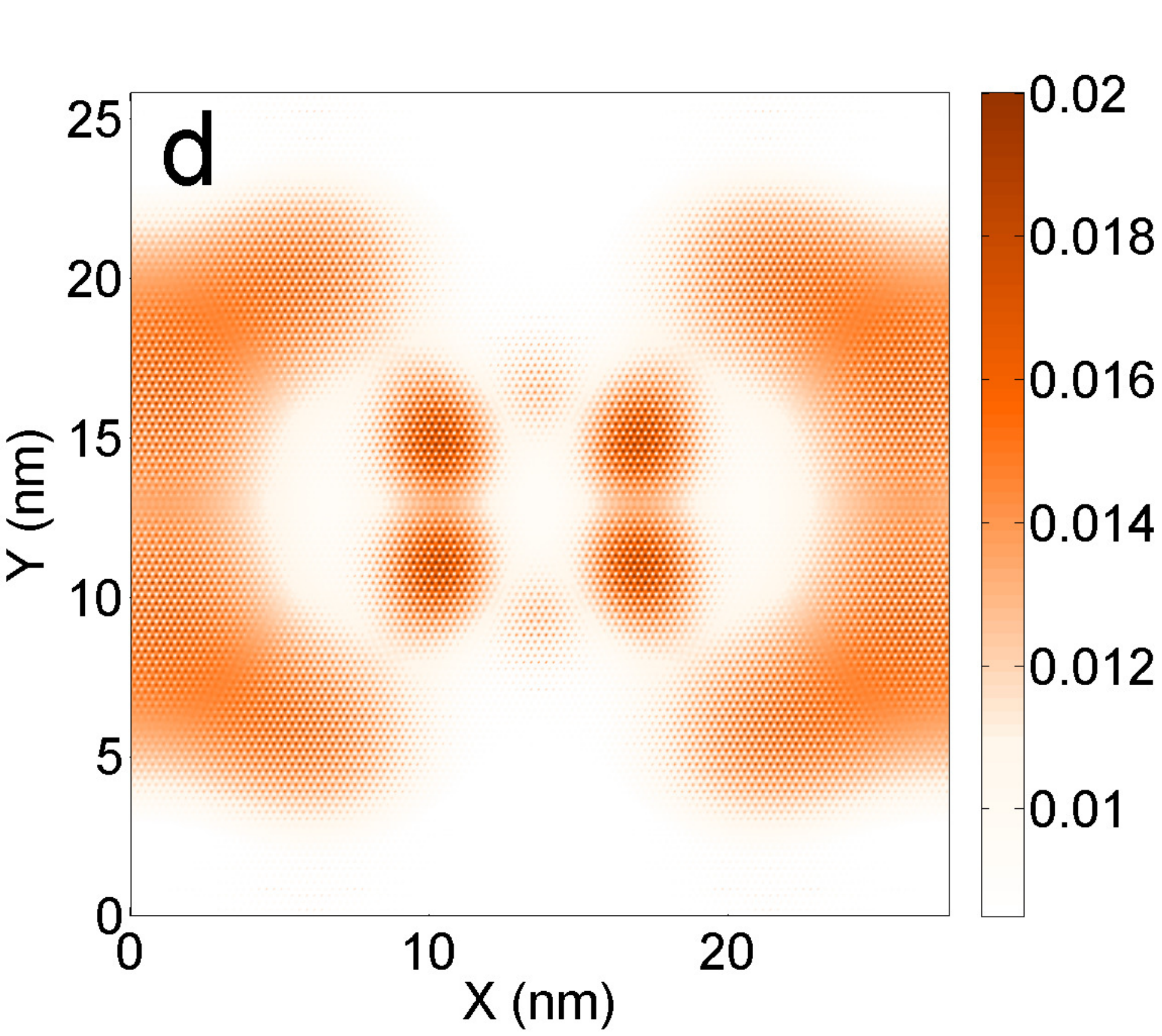}
\caption{(Color online) LDOS for: AGNR [panels a), b)] and ZGNR
[panels c) and d)] with deformation amplitude $A=1.42$ nm and
$b=4.3$ nm at energies shown in Fig. \ref{Fig2} (vertical dashed
lines): a) (blue) $E=0.1$ eV, b) (red) $E=0.15$ eV, c) (green)
$E=0.15$ eV and d) (orange) $E=0.21$ eV. Scales (a.u.) in each
plot are optimized to exhibit areas with higher density of
states.} \label{Fig4}
\end{figure}
\begin{figure}
\includegraphics[scale=0.34]{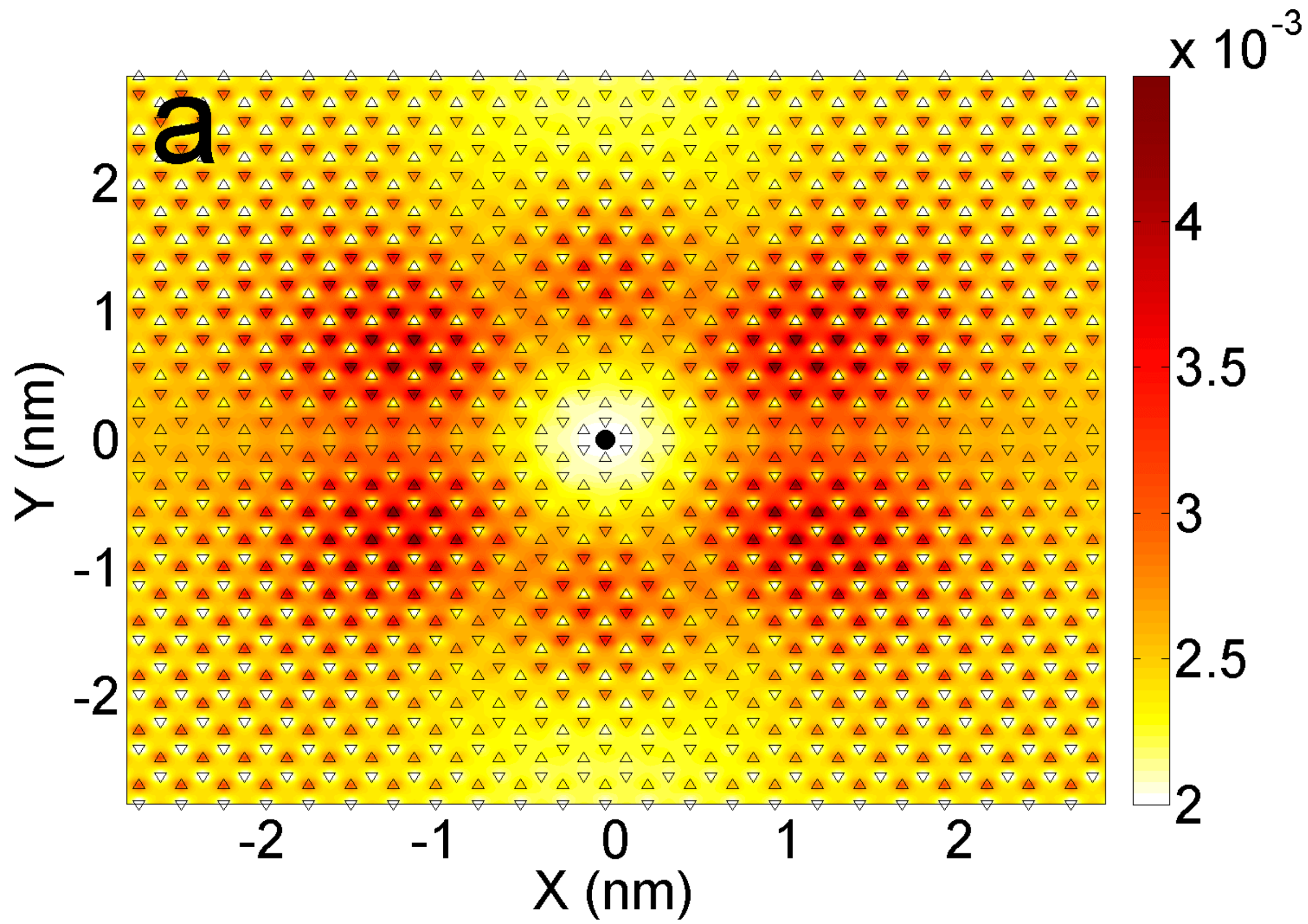}
\includegraphics[scale=0.205]{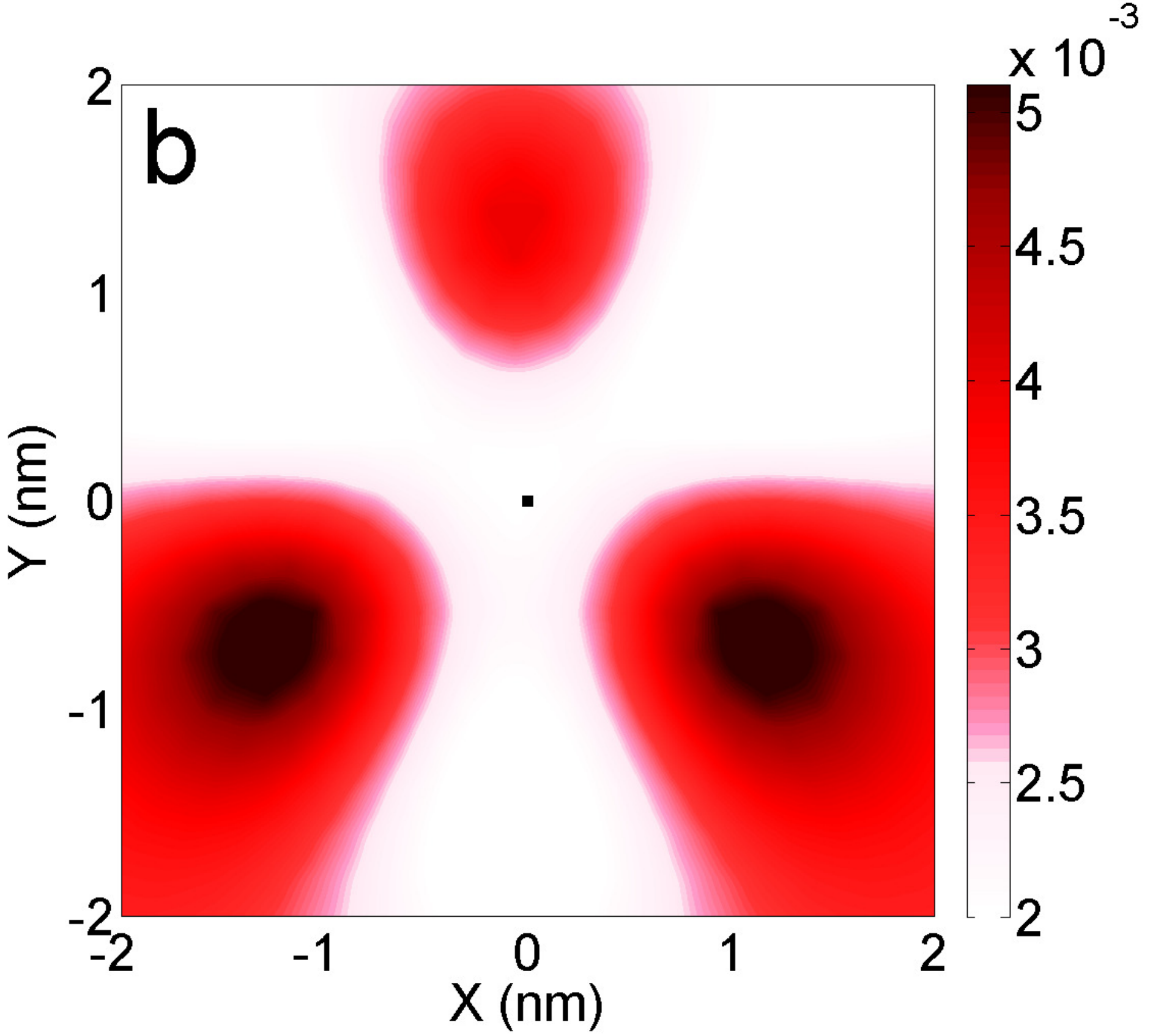}
\includegraphics[scale=0.205]{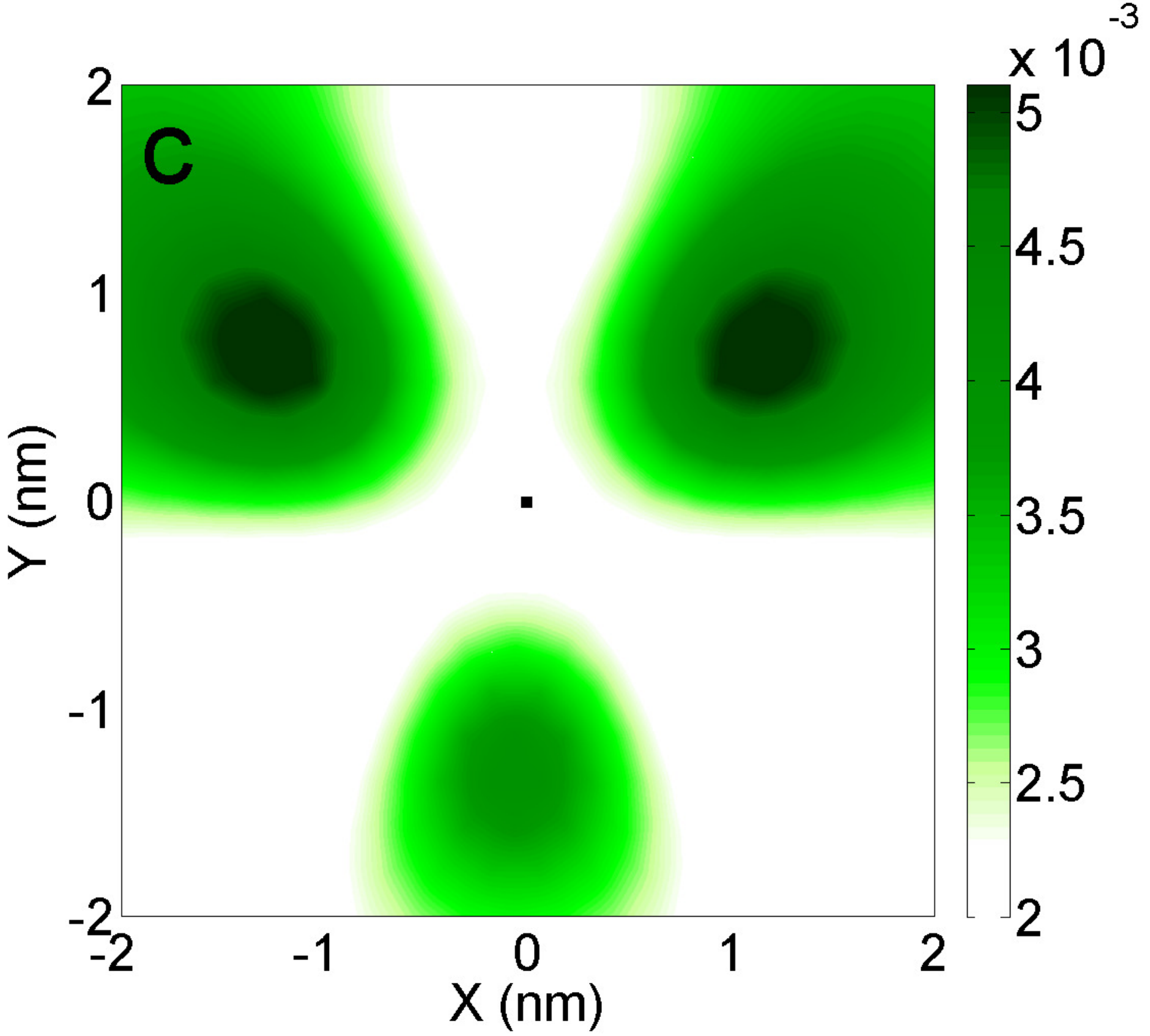}
\caption {(Color online) a) LDOS (a.u.) for ZGNR of width $W=25.8$
nm, and length $L=27.3$ nm, with Gaussian deformation of amplitude
$A=0.7$ nm, and $b=1.4$ nm, plotted at energy $E=0.042$ eV. Atomic
positions  of undeformed graphene lattice marked with up- and
downside triangles for each sublattice. The center of the Gaussian
is marked with the black circle. b) Sublattice A. c) Sublattice B.
\label{Fig5}}
\end{figure}

\noindent{\it LDOS and Pseudo-Spin Polarization}: Non-homogeneous
strain has profound effects on the space distribution of the
DOS. An analysis of the LDOS reveals a well-defined
pattern with a $60^{\circ}$ symmetry, i.e., the 'petals' of the
'flower' structure. Fig.~\ref{Fig4} presents typical LDOS
structures obtained for AGRN and ZGNR at energies marked by
vertical lines in Fig.~\ref{Fig2}. We have confirmed that this
structure persists for a wide energy range and deformation
parameter values (not shown). Similar patterns have been obtained
in models for closed
systems\cite{Wakker,Barraza1,Barraza2,Barraza3,Moldovan}. Notice
that the structures for ZGNR and AGNR are rotated $90^{\circ}$ relative to each other,
following the spatial distribution of the
pseudo-magnetic field\cite{Daiara}.

Fig.~\ref{Fig5} shows a zoom-in of one particular structure, for a
ZGNR. The undeformed graphene lattice is represented by up- and
downside triangles (distinguishing sublattices). The black dot
represents the maximum height of the Gaussian bump that is
centered in a maximum symmetry position in the ribbon. The values
for sublattice occupancy alternates from 'petal' to 'petal',
signaling a characteristic sublattice asymmetry or pseudo spin
polarization with 3-fold symmetry. Such structure could be linked
to a geometrical description of the microscopic model as realized
in Ref.~\onlinecite{Barraza3}. Similar effects, with chiral
states within the zero LL\cite{Goerbig}, were obtained in models of Dirac fermions
with magnetic field in bounded regions\cite{deMartino}.

In panels b) and c) we show values of LDOS on each sublattice.
Panel b) exhibits the largest occupancies (darker regions) at the
bottom 'petals', while the contrary occurs in panel c). Notice
that zigzag boundaries naturally introduce a difference in
sublattice occupancies due to the different sublattice
terminations at the top and bottom edges. These differences, due
to 'edge states', are predicted to be localized at the edges,
however for finite systems the amplitude of edge states decays
inside the ribbon\cite{Fujita}. It is thus natural to interpret
the 'darker regions' breaking the three-petal symmetry as a
consequence of edge states in ZGNR. To confirm these hypothesis we
carried out calculations for AGNR that reveal the same alternate
pattern for sublattice occupancy\cite{Supplementary}. In these
systems the whole 'flower' structure possesses 'dark regions'
in the 'petals' appearing closer to the
contacts to reservoirs. Although the leads are modeled as perfect
graphene lattices, the absence of the deformation in the
reservoirs could create an effective boundary condition at the
contact, thus representing potentially a zigzag boundary.  The
presence of these developing 'edge states' at the contacts could
enhance the sublattice occupancy in certain petals. Calculations
carried out in larger AGRN with deformation amplitudes vanishing
before reaching contact regions (thus eliminating a 'zigzag
boundary'), show that the distribution of the highest occupied
'petals' becomes energy dependent with a persistence asymmetry between petals.
However this asymmetry decreases with increasing AGNR width, suggesting a
strong dependence on the underlying LDOS for the undeformed system.

Further calculations reveal that pseudo spin polarization appears
in a wide range of energies, and deformation parameters,
indicating a robust effect, that persists in the presence of
external magnetic fields\cite{Martin}. Note that this local
breaking of sub lattice symmetry (local breaking of inversion symmetry) does not open a gap as evidenced by the finite conductance. Although several theory
studies have predicted sublattice asymmetry features in the
LDOS\cite{Moldovan,Neek-Amal,Barraza2,Barraza3}, these appear to
have been overlooked in STM experimental studies
\cite{Xu,Morgenstern,Burke} since no explicit connection with
centro-symmetric deformations have been made. Our results, showing
a peculiar sequential pattern for sublattice occupation provide a
possible test for the origin of the  observed asymmetries that
could be tested in current experimental settings\cite{Alex2}.

\noindent{\it Conclusions:} In closing, we present the first study of conductance of strained ribbons with Gaussian deformations that produce inhomogeneous pseudo magnetic fields at every length scale. In this system there are no Landau levels available for transport but instead there are bound states that concentrate in the region where the pseudo magnetic field acquires its maximum value. We provide a real space description of the location and symmetry of these states, that exhibit a sublattice occupation alternation
of $60^{\circ}$, associated with a local pseudo spin polarization in the
continuum Dirac (low-energy) description. These results are largely independent of lattice orientation. All these effects are within reach of current experiments and open the exciting
possibility to design deformations for desired electronic
confinement.

\noindent{\it Acknowledgments } We acknowledge
discussions with M. Schneider, S. V. Kusminskiy, M. Morgenstern,
A. Georgi, S. Ulloa, G. Petersen, M. M. Asmar and F. de Juan.
R.C-B., D.F and N.S. acknowledge the hospitality of the Dahlem
Center for Quantum Complex Systems at Freie Universitat where this
project was initiated. This work was supported by NSF No.
DMR-1108285 (D.F., R.C-B. and N.S.), FAPERJ E-26/101.522/2010
(A.L.); CNPq, CAPES (2412110) and DAAD (D.F.);
CONACYT, PAPIIT-DGAPA UNAM IN109911 (R.C-B, F.M.).
\bibliography{Refs}
\end{document}